\documentclass[12pt,aps,prb,preprint,showpacs,showkeys]{revtex4}  

\usepackage{amsmath}    
\usepackage{amsfonts}   
\usepackage{amssymb}
\usepackage{graphicx}   
\usepackage{subfigure}

\begin{document}
\baselineskip = 7.5mm \topsep=1mm
\begin{center} {\LARGE\bf ``Chiral" universality class behavior of a non-chiral antiferroquadrupole system} \vspace{5 mm}\\ M. \v{Z}ukovi\v{c}$^{\rm a*}$, T. Idogaki$^{\rm b}$, and K.Takeda$^{\rm a,b}$
\vspace{5 mm}\\
\end{center}
\noindent $^{\rm a}$ Institute of Environmental Systems, Kyushu University, Fukuoka 812-8581, Japan
\newline
$^{\rm b}$ Department of Applied Quantum Physics, Kyushu University, Fukuoka 812-8581, Japan
\vspace{10mm}
\newline
The planar Heisenberg system with antiferroquadrupolar exchange on 3D stacked triangular lattice is
shown to belong to the new ``chiral" universality class, predicted for chiral antiferromagnets. The
present system, however, displays no such chirality, which, according to the currently widely accepted
concept, is supposed to be a key ingredient for the new critical behavior. Our claim that the new
universality class should not be limited only to the chiral antiferromagnets is based on a simple
mapping between the chiral antiferromagnetic and antiferroquadrupolar systems and supported by actual
calculations of the transition temperatures and critical indices from finite-size scaling analysis of
data issued from Histogram Monte Carlo simulations. In order to demonstrate the existence of two
different universality classes in the behavior of quadrupolar systems, the finite-size scaling is also
performed for the system with ferroquadrupolar exchange, which, in contrast to the antiferroquadrupolar
exchange system case, produced standard critical behavior, as could also be anticipated from the
mapping. \vspace{10 mm}
\\ $PACS\ numbers$: 75.10.Hk; 75.30.Kz; 75.40.Cx; 75.40.Mg.\vspace{10 mm}
\newline
$*$Author to whom correspondence should be addressed. Permanent address: Institute of Environmental
Systems, Faculty of Engineering, Kyushu University, Higashi-ku, Hakozaki 6-10-1, Fukuoka 812-8581,
Japan. Fax: +81-92-633-6958.

\newpage
\indent It has been theoretically predicted \cite{kawamura1,kawamura2} and supported by experiment
\cite{ajiro_etal}-\cite{deutschmann_etal} that geometrically frustrated XY antiferromagnets on a stacked
triangular lattice (STL), described by Hamiltonian
\begin{equation}
\label{H} {\mathcal{H}}= -J\sum_{\langle i,j \rangle}\mbox{\boldmath $S$}_{i} \cdot \mbox{\boldmath
$S$}_{j} \ ,
\end{equation}
where $\mbox{\boldmath $S$}_{i} = (S_{ix},S_{iy})$ is a two-dimensional unit vector at the $i$th lattice
site , $\langle i,j \rangle$ denotes the sum over nearest neighbors, and $J<0$ is the bilinear exchange
interaction constant, should belong to a new universality class linked to the two-fold chiral degeneracy
inherent to the $120^{\circ}$ ordered spin structure. Critical behavior of these systems is
characterized by new critical indices, different from those for non-frustrated systems with XY spin
symmetry.
\newline
\indent Motivated by the existence of two different universality classes in the critical behavior of the
geometrically frustrated chiral antiferromagnets and non-frustrated systems with XY spin symmetry, we
investigated critical behavior of the corresponding systems with only biquadratic exchange interaction.
Namely, we studied models described by the Hamiltonian
\begin{equation}
\label{H'} {\mathcal{H'}}= -J'\sum_{\langle i,j \rangle}(\mbox{\boldmath $S$}_{i} \cdot \mbox{\boldmath
$S$}_{j})^{2} \ ,
\end{equation}
where $J'$ is the biquadratic exchange interaction constant, in which we considered both $J'>0$ and
$J'<0$. For the case of $J'>0$ (ferroquadrupolar exchange), the system is non-frustrated and is expected
to belong to the standard 3D XY universality class. On the other hand, the system with $J'<0$
(antiferroquadrupolar exchange) is frustrated due to the triangular lattice geometry, resulting in the
non-collinear ground state (Fig.1). However, such a system, compared to the one with the
antiferromagnetic exchange (Fig.2), lacks the chirality - an essential element which is supposed to be
responsible for the new critical behavior of the STL antiferromagnets. The question is: Will such a
system still display the standard universality class behavior?
\newline
\indent Since the universality class is in principle determined by the space dimensionality and symmetry
of the order parameter, let us first consider the issue of the order parameter symmetry in the cases of
the STL antiferromagnet described by Hamiltonian (\ref{H}) and the STL antiferroquadrupolar system
described by Hamiltonian (\ref{H'}) (the space dimensionality is the same). XY continuous symmetry is
common for both systems, however, there is an additional ingredient which makes the two systems
different. In the case of the STL antiferromagnet one may define a local chirality at each elementary
triangular plaquette, by
\begin{equation}
\label{chirality} \kappa = \frac{2}{3\sqrt{3}}\sum_{\langle i,j \rangle}^{p}[\mbox{\boldmath $S$}_{i}
\times \mbox{\boldmath $S$}_{j}]_{z} \ ,
\end{equation}
where the summation runs over the three directed bonds surrounding each plaquette, $p$, which is an
Ising-like quantity representing the sign of rotation of the spins along the three sides of each
plaquette. Due to the chirality the STL antiferromagnet has two-fold degeneracy of the ground state
($\kappa=+1$ and $\kappa=-1$), resulting in the structure with spins arranged on plaquettes with turn
angles $+120^{\circ}$ and $-120^{\circ}$, respectively. A minimum energy condition is realized by an
arrangement in which the $+$ and $-$ plaquettes alternate, as shown in Fig.2, producing long-range
chiral order at low temperatures. On the other hand, as far as the chirality is concerned, the STL
antiferroquadrupolar system has four-fold degeneracy in the ground state of each plaquette ($\kappa=\pm
1,\pm \frac{1}{3}$), resulting in the structure with four possible turn angles between two neighbouring
spins $\pm 120^{\circ}$, $\pm 60^{\circ}$ (Fig.1). However, unlike in the chiral case, there is no
energetically favorable arrangement among the four kinds of plaquettes in the antiferroquadrupolar
system and, hence, the plaquettes do not order even at low temperatures. So, compared to the chiral
antiferromagnet with both Ising and rotational symmetries of the order parameter (Z$_{2}$ $\times$
O($n$=2)), the order parameter symmetry of the system with antiferroquadrupolar exchange is apparently
different. In the following we will show, by a simple mapping between the dipolar and quadrupolar forces
\cite{carmesin}, that this difference, however, should not constitute grounds for different critical
behavior.
\newline
\indent The Hamiltonians ${\mathcal{H}}$ and ${\mathcal{H'}}$ can be rewritten in the respective forms
\begin{equation}
\label{H1} {\mathcal{H}}= -J\sum_{\langle i,j \rangle}\cos(\varphi_{i}-\varphi_{j})\ ,
\end{equation}
and
\begin{equation}
\label{H'1} {\mathcal{H'}}= -J'\sum_{\langle i,j \rangle}\cos^{2}(\varphi_{i}-\varphi_{j})\ ,
\end{equation}
where $\varphi_{i}-\varphi_{j}$ is an angle between the neighbouring spins $\mbox{\boldmath $S$}_{i}$
and $\mbox{\boldmath $S$}_{j}$. Using the identity $\cos^{2}(\alpha)=(1+\cos2\alpha)/2$ we can write the
corresponding partition functions in the forms
\begin{equation}
\label{Z} {\mathcal{Z}}(J,T)=
\int_{0}^{2\pi}\prod_{k=1}^{N}\mathrm{d}\varphi_{k}\exp\left(\frac{J}{T}\sum_{\langle i,j
\rangle}\cos(\varphi_{i}-\varphi_{j})\right)\ ,
\end{equation}
and
\begin{equation}
\label{Z'} {\mathcal{Z'}}(J',T)= \left(2\pi \exp\frac{2J'}{T}\right)^{N}
\int_{0}^{2\pi}\prod_{k=1}^{N}\mathrm{d}\varphi_{k}\exp\left(\frac{J'}{2T}\sum_{\langle i,j
\rangle}\cos2(\varphi_{i}-\varphi_{j})\right)\ ,
\end{equation}
It can easily be seen that if we substitute $\vartheta_{i}-\vartheta_{j}=2(\varphi_{i}-\varphi_{j})$
into Eqn. (\ref{Z'}) and assume $|J|=|J'|$, we have
\begin{equation}
\label{ZZ'} {\mathcal{Z}}(J,T) = \left(2\pi \exp\frac{2J'}{T}\right)^{-N}{\mathcal{Z'}}(J',T/2) \ ,
\end{equation}
which holds for the case of a non-frustrated systems with collinear ordering, as well as the present
frustrated systems case with non-collinear ordering. Therefore, in either case we can expect the same
critical behavior of the systems (\ref{H}) and (\ref{H'}) at the respective transition temperatures
$T_{c}$ and $T_{q}=T_{c}/2$.
\newline
\indent In order to confirm the prediction based on the mapping, we further performed a Histogram Monte
Carlo (HMC) simulation \cite{ferr-swen} analysis. We considered the systems of $N=L^{3}$ spins, where
the lattices linear dimension $L$ = 12, 18, 24 and 30, assuming periodic boundary condition throughout.
Spin updating followed the Metropolis dynamics and averages were calculated using $2\times10^{6}$ Monte
Carlo steps (MCS) after discarding another $1\times10^{6}$ MCS for thermalization. Then we used the data
issued from the HMC simulation in order to perform a finite-size scaling analysis of the following
physical quantities: the quadrupole long-range order (QLRO) parameter $q$,
\begin{equation}
\label{eq.q}q=\langle Q \rangle/N,\ {\mathrm{where}}\
Q=\left[\left(\sum_{i}\left(\left(S_{ix}\right)^{2}-\left(S_{iy}\right)^{2}\right)\right)^{2}+
\left(\sum_{i}2S_{ix}S_{iy}\right)^{2}\right]^{\frac{1}{2}}\ ,
\end{equation}
the corresponding susceptibility per site $\chi_{q}$
\begin{equation}
\label{eq.chi}\chi_{q} = \frac{(\langle Q^{2} \rangle - \langle Q \rangle^{2})}{Nk_{B}T}\ ,
\end{equation}
and the logarithmic derivatives of $\langle Q \rangle$ and $\langle Q^{2} \rangle$ with respect to
$K=1/k_{B}T$
\begin{equation}
\label{eq.D1}D_{1q} = \frac{\partial}{\partial K}\ln\langle Q \rangle = \frac{\langle QE
\rangle}{\langle Q \rangle}- \langle E \rangle\ ,
\end{equation}
\begin{equation}
\label{eq.D2}D_{2q} = \frac{\partial}{\partial K}\ln\langle Q^{2} \rangle = \frac{\langle Q^{2} E
\rangle}{\langle Q^{2} \rangle}- \langle E \rangle\ .
\end{equation}
Temperature-dependences of these quantities display extrema at the $L$-dependent transition temperatures
$T_{q}(L)$, which at a second-order transition are known to scale with a lattice size as:
\begin{equation}
\label{eq.scalchi}\chi_{q,max}(L) \propto L^{\gamma_{q}/\nu_{q}}\ ,
\end{equation}
\begin{equation}
\label{eq.scalV1}D_{1q,max}(L) \propto L^{1/\nu_{q}}\ ,
\end{equation}
\begin{equation}
\label{eq.scalV2}D_{2q,max}(L) \propto L^{1/\nu_{q}}\ ,
\end{equation}
\noindent where $\nu_{q}$ and $\gamma_{q}$ represent the correlation length and susceptibility critical
indices, respectively.
\newline
\indent In Fig.3 we present the scaling analysis for the case of $J'>0$ in ln-ln plot. As expected, the
transition is of second order and the obtained values of the critical indices, $\nu_{q}$ = 0.667
$\pm$0.008 and $\gamma_{q}$ = 1.333 $\pm$ 0.018, are close to those for the standard XY universality
class (see Table 1). Similar scaling analysis was done for the system with the antiferroquadrupolar
couplings ($J'<0$) and is presented in Fig.4. The transition is again of second order, however, the
values of the critical indices $\nu_{q}$ =0.520 $\pm$ 0.003 and $\gamma_{q}$ = 1.072 $\pm$ 0.009 turn
out to be quite different from the standard ones but strikingly close to those of the new chiral
universality class. We note that for the sake of comparison we also performed scaling analysis for the
antiferromagnetic system (\ref{H}) with $J<0$ and the calculated values of the critical indices, $\nu$ =
0.514 $\pm$ 0.007 and $\gamma$ = 1.074 $\pm$ 0.013, (see Table 1 for comparison with the values obtained
from some other studies) were indeed found to be in a nice agreement with those for the
antiferroquadrupolar system. The comparison of the critical temperatures $T_{c}$ and $T_{q}$ also shows
that they indeed differ by the factor two and, hence, along with the critical indices (all summarized in
Table 2) excellently corroborate the prediction based on the mapping. Therefore, the present result
would suggest that not necessarily the chirality as defined in equation (\ref{chirality}) but rather
frustration, resulting in the non-collinear ground state, is the element causing the novel critical
behavior.
\newline
\indent In summary, the results presented in this work provided support for existence of the new
universality class predicted by Kawamura \cite{kawamura1,kawamura2}. However, the completely new result
is that the system with antiferroquadrupolar exchange on STL, which has no chirality possessed by its
counterpart with antiferromagnetic exchange, was found to display critical behavior strikingly similar
to that of the chiral STL antiferromagnet of the new universality class. Based on the mapping between
the two systems and HMC numerical results, we conclude that the two systems should indeed belong to the
same universality class, that so far has been thought to be limited only to the chiral antiferromagnets.

\newpage

\vspace*{80mm}

\begin{table}[h]
\caption{Critical indices $\nu,\ \gamma$ for dipole ordering - previous studies.} \label{tab.y}
\begin{center}
\begin{tabular}{|c||c|c|c|}                                                                \hline
                          & $\nu$           & $\gamma$        & Ref.   \\\hline \hline
 Standard XY model class  & 0.669           & 1.316           & \cite{guillou}    \\ \hline \hline
 Chiral XY model class    & 0.54 $\pm$ 0.02 & 1.13 $\pm$ 0.05 & \cite{kawamura1}     \\ \cline{2-4}
                          & 0.57 $\pm$ 0.03 & 1.10 $\pm$ 0.05 & \cite{ajiro_etal}    \\ \cline{2-4}
                          & 0.54 $\pm$ 0.03 & 1.01 $\pm$ 0.08 & \cite{mason_etal}    \\ \hline

\end{tabular}
\end{center}
\end{table}

\newpage

\vspace*{80mm}

\begin{table}[h]
\caption{Critical indices $\nu_{q},\ \gamma_{q}$ and $\nu,\ \gamma$ for quadrupole and dipole ordering,
respectively.} \label{tab.x}
\begin{center}
\begin{tabular}{|c||c|c|c|}                                                                \hline
         & $\nu_{q}$           & $\gamma_{q}$       & $T_{q}/|J'|$      \\ \hline \hline
 $J'>0$  & 0.667 $\pm$ 0.008   & 1.333 $\pm$ 0.018  & 1.520 $\pm$ 0.002  \\ \hline
 $J'<0^{\dag}$  & 0.520 $\pm$ 0.003   & 1.072 $\pm$ 0.009  & 0.729 $\pm$ 0.002  \\ \hline
         & $\nu$               & $\gamma$           & $T_{c}/|J|$      \\ \hline \hline
 $J<0$     & 0.514 $\pm$ 0.007   & 1.074 $\pm$ 0.013    & 1.458 $\pm$ 0.002         \\ \hline
\end{tabular}
\end{center}
$^{\dag}$ Since the sign of the inter-plane interaction $J'_{z}$ is irrelevant in this case, for
simplicity we used $J'_{z}>0$.
\end{table}

\newpage

\begin{figure}[!t]
\includegraphics[scale=0.5]{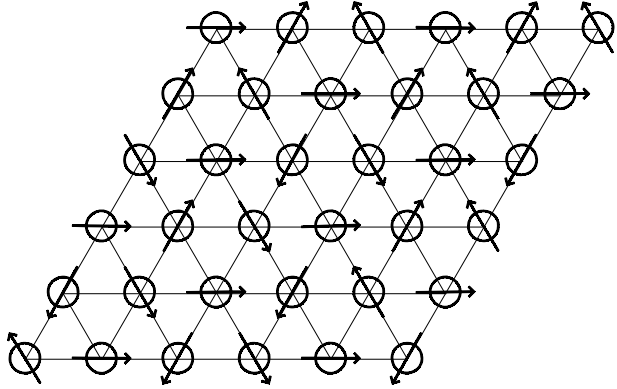}
\caption{Ground state spin configuration of the system with antiferroquadrupolar
exchange ($J'<0$) shown in one triangular plane. In contrast to the frustrated XY antiferromagnets
(Fig.2), the elementary triangles do not show chiral ordering.}
\end{figure}

\begin{figure}[!t]
\includegraphics[scale=0.5]{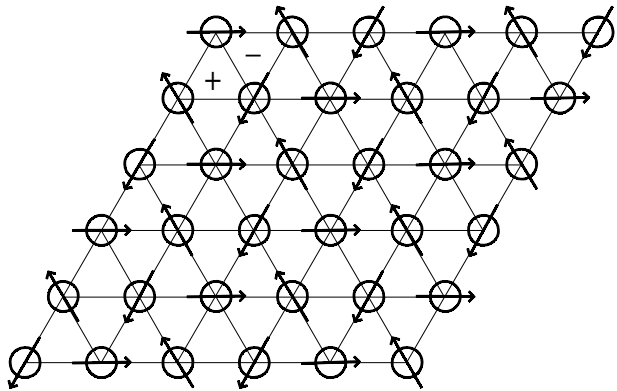}
\caption{Ground state spin configuration of the system with antiferromagnetic
exchange ($J<0$). The ``up" ($+$) and ``down" ($-$) triangles consisting of spins with turn angles
$120^{\circ}$ and $-120^{\circ}$, respectively, show chiral ordering.}
\end{figure}

\begin{figure}[!t]
\includegraphics[scale=0.5]{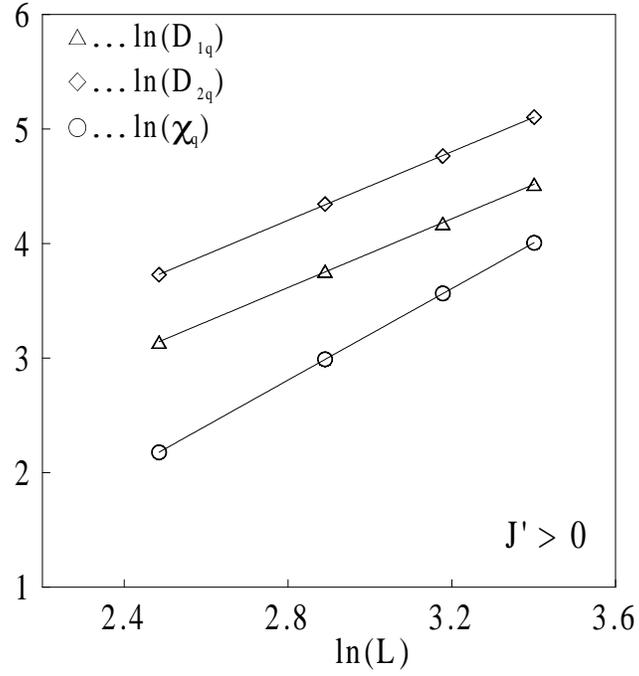}
\caption{Scaling of maxima of the quantities $\chi_{q}$, $D_{1q}$ and
$D_{2q}$ in ln-ln plot for the case of $J'>0$. The slopes yield values of $1/\nu_{q}$ for $D_{1q},\
D_{2q}$ and $\gamma_{q}/\nu_{q}$ for $\chi_{q}$.}
\end{figure}

\begin{figure}[!t]
\includegraphics[scale=0.5]{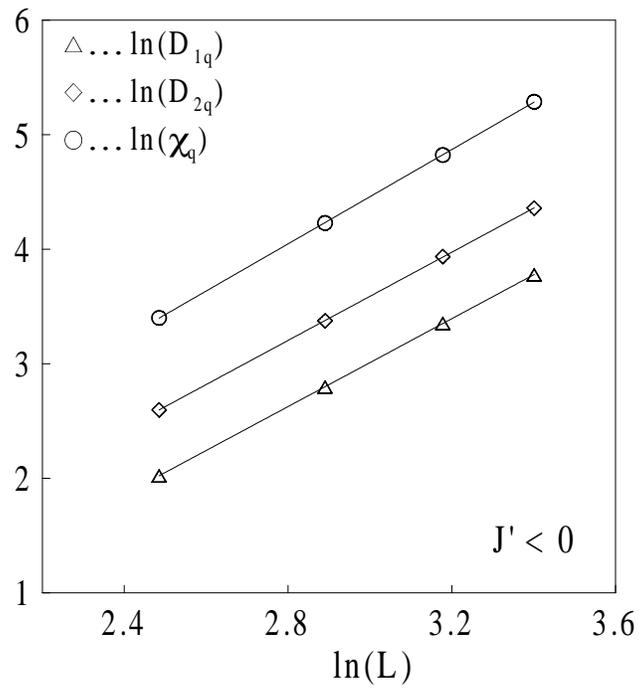}
\caption{The same as in Fig.3 for the case of $J'<0$.}
\end{figure}

\end{document}